\documentclass[9pt,twocolumn,twoside]{opticajnl}
\journal{opticajournal} 

\setboolean{shortarticle}{true}

\usepackage{lineno}

\usepackage{subcaption}
\usepackage{mathtools}

\DeclarePairedDelimiterX\braket[2]{\langle}{\rangle}{#1\,\delimsize\vert\,\mathopen{}#2}
\DeclarePairedDelimiterX\ketbra[2]{\lvert}{\rvert}{#1\rangle\hspace*{-\nulldelimiterspace}\mathopen{}\langle#2}

\definecolor{darkspringgreen}{rgb}{0.09, 0.45, 0.27}

\title{Propagation-Invariant Transmission Eigenmodes in Weakly Scattering Media}

\author[1]{J. C. Osorio--Tascón}
\author[2]{P. A. Quinto--Su}

\affil[ ]{Instituto de Ciencias Nucleares, Universidad Nacional Autónoma de México, Apartado Postal 70-543, 04510, Cd. Mx., Mexico}

\affil[1]{osoriotascon.jc@gmail.com}
\affil[2]{pedro.quinto@nucleares.unam.mx}

\begin{abstract}
	Propagation-invariant beams (PIBs) with annular angular spectra preserve their transverse intensity structure over extended propagation distances. Here, we investigate whether scattering invariant modes (SIMs) can simultaneously exhibit scattering and propagation invariance. We measure free-space and sample transmission matrices (TMs) in an orthonormal annular basis for weak, moderate, and strong scattering media and identify localized SIMs (LSIMs) whose angular spectra remain predominantly confined to one or a few neighboring rings. Under weak scattering, $75$ of the $128$ SIMs considered for annular localization satisfy this criterion, with LSIMs exhibiting Bessel-like propagation. Under moderate scattering, $25$ LSIMs are recovered after filtering low-intensity spectral components, whereas none are found under strong scattering. Although increasing modal crosstalk progressively suppresses annular localization and nondiffracting behavior, SIMs retain higher transverse-intensity correlation over propagation than the Gaussian-like and Bessel beams in the moderate and strong regimes. These results show that annular localization enables SIMs to acquire propagation invariance when scattering is weak, whereas the enhanced propagation robustness of SIMs can persist even after this localization is lost.
\end{abstract}

\setboolean{displaycopyright}{false} 

\begin{document}
	
	\maketitle
	
	Light transmitted through random media undergoes multiple scattering, producing speckle patterns that can severely degrade the incident spatial mode and polarization. This degradation limits applications like imaging through turbid media, optical focusing, and information transmission. To overcome these effects, TM approaches have emerged as powerful tools for controlling light propagation through random media \cite{Popoff2010}, coupling light into high-transmission eigenchannels \cite{Kim2012}, reconstructing complex optical fields \cite{Liu2019}, and compensating aberrations in deep-tissue imaging \cite{Badon2020}. Alternative strategies seek optical fields that are intrinsically robust against scattering. For example, structured vector beams can preserve specific polarization distributions after transmission through random media \cite{Nape2022}, while topological optical fields, such as skyrmions, have also demonstrated enhanced resilience \cite{Peters2025}. PIBs provide another promising class because their self-healing and nondiffracting properties make them attractive for imaging and beam delivery through complex environments.
	
	Many applications, particularly those involving dynamic scattering media such as biological tissue or atmospheric turbulence, require rapid characterization of the transmitted wavefront of the TM before the medium changes. One approach to reduce the number of required measurements is via compressive sensing \cite{Scheidt2023hadamard}. Alternatively, measuring the TM in an annular angular spectrum basis can reduce the number of spatial modes for wavefront optimization relative to a conventional representation \cite{Scheidt2023}. This last approach has the additional advantage that it is the natural basis to generate PIBs \cite{Mahmoud2013}.
	
	Recently, Pai et al. \cite{Pai2021} introduced the concept of SIMs, TM eigenmodes that produce identical transmitted fields whether they propagate through a scattering medium or free-space. Owing to this property, SIMs have shown promise for coherent imaging in situations where ballistic light is scarce \cite{Badon2020} and as robust carriers of optical information \cite{Singh2023}. However, conventional SIMs are not propagation-invariant. Their transverse profiles diffract during free-space propagation in a manner similar to Gaussian beams, resulting in a limited depth of focus (DOF) and a rapid decrease in field correlation away from the observation plane.
	
	This raises a natural question: can SIMs also be made propagation-invariant? We address this question by measuring the TM in an annular angular spectrum basis and examining whether the resulting SIMs can simultaneously retain scattering invariance and the propagation invariance associated with annular spectra. To assess the role of modal crosstalk, we perform measurements with three samples spanning weak, moderate, and strong scattering regimes. Finally, we analyze the angular spectrum structure and propagation characteristics of the resulting SIMs and compare their performance with Gaussian-like and Bessel beams.

	\begin{figure}
		\centering
		\includegraphics[width=\linewidth]{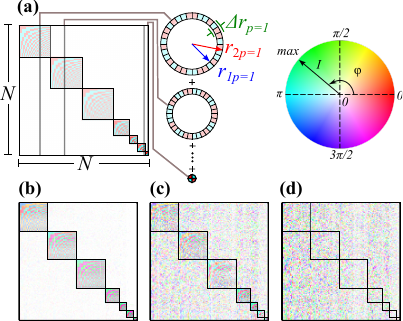}
		\caption{(a) Input basis of $N=132$ modes with $P=6$ concentric rings of width $\Delta r_p = \left[0.48, 0.46, 0.46, 0.36, 0.34, 0.24\right]$~mm, inner and outer radius $r_{1p}$ and $r_{2p}$, respectively. (b-d) Measured $T_\text{s}$ with the (b) weak, (c) moderate and (d) strong scattering samples. The ratio between the average off-diagonal and diagonal intensities are (b) $(5.56\pm 0.16)\times 10^{-4}$ (c) $(1.58\pm 2.26)\times 10^{-2}$ and (d) $0.69\pm 0.99$. Here the color scheme represents intensity ($I$) and phase ($\varphi$).}
		\label{fig: 1}
	\end{figure}
	
	\begin{figure*}[!t]
		\centering
		\includegraphics[width = 0.7\textwidth]{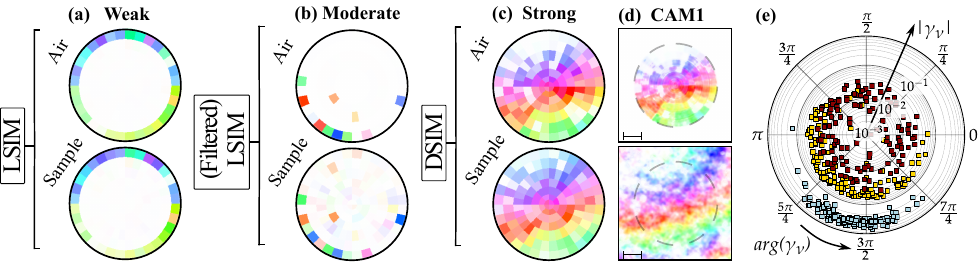}
		\caption{(a-c) Experimental SIMs represented in the annular basis after propagating in air and the sample. (d) The entire field of (c) recorded by CAM1; the dashed-line circumference represents the maximum aperture covered by the basis. (e) The log--polar plot of the measured SIM eigenvalue $\gamma_\nu$. Here the color scheme is the same as in Fig.~\ref{fig: 1}. The bar scale in (c) is $300~\mu$m.}
		\label{fig: 2}
	\end{figure*}
	
	\begin{figure*}[!t]
		\centering
		\includegraphics[width = \textwidth]{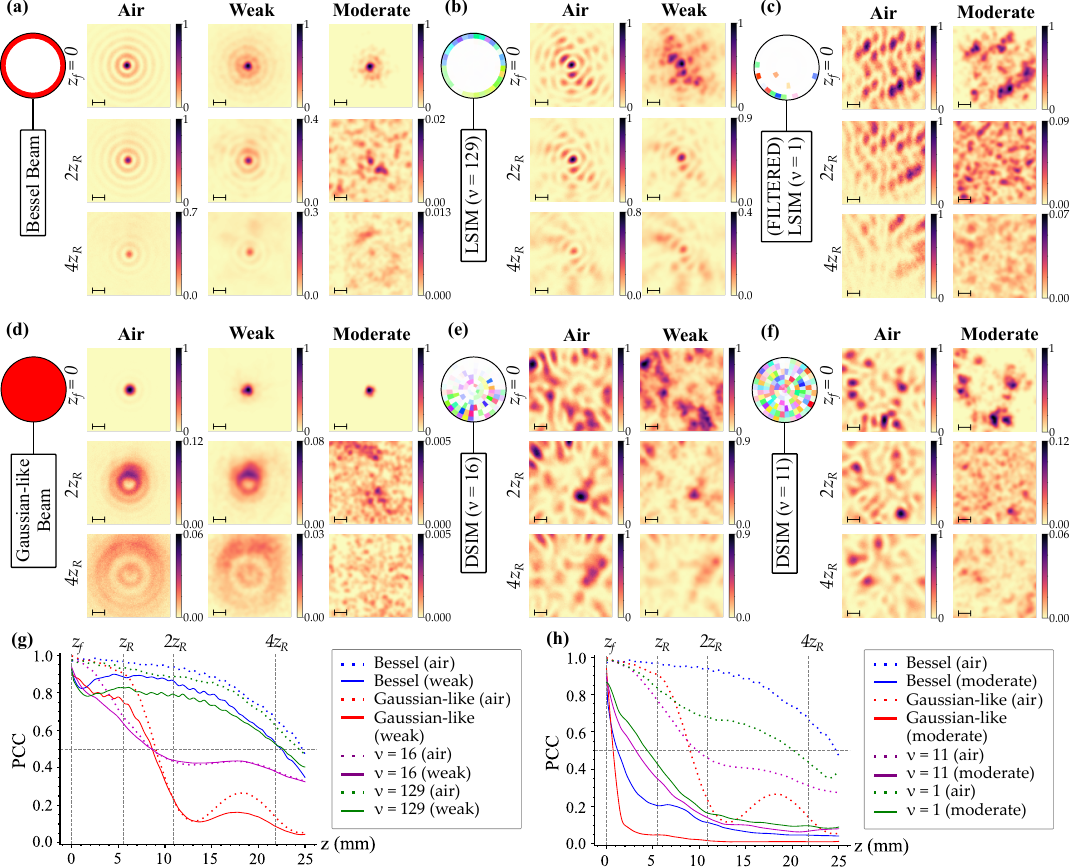}
		\caption{
			(a--f) Intensity evolution at planes $z_f=0$, $2z_R$ and $4z_R$ in the weak and moderate scattering samples for the (a) Bessel beam, (b-c) LSIMs, (d) Gaussian-like beam and (e-f) DSIMs. (Left insets) Input field. (g-h) The correlation (PCC) calculated for the propagated fields in (dash--dotted lines) air and (solid lines) both (g) weak and (h) moderate scattering samples. The scale bar is $50~\mu$m.
		}
		\label{fig: 3}
	\end{figure*}
	
	A HeNe laser operating at a wavelength of $\lambda_0=632.8$~nm illuminates a phase-only spatial light modulator (SLM), which defines the input optical field in the angular spectrum ($k$-space) domain (see Fig.~S1). A 4$f$ imaging system relays the SLM plane to the output plane, where the complex transmitted field is recorded by CAM1. The random scattering sample is placed at the Fourier plane of the first lens, such that the measured TM describes the transformation between the incident angular spectrum and the transmitted field. The complex field is recovered using three-step phase-shifting interferometry (PSI) by interfering the transmitted field with a reference beam.
	To characterize propagation, a secondary imaging arm focuses the transmitted field onto CAM2, which is mounted on a motorized translation stage with a travel range of 25 mm. Recording the transverse intensity distribution at successive axial positions provides the propagation evolution of the generated fields. Three Thorlabs ground-glass diffusers (1500, 600, and 220 grit) are used as random scattering media and are referred here as weak, moderate and strong scattering media, respectively, according to the amount of modal crosstalk they introduce into the measured TM.
	
	
	Since PIB possess annular angular spectra, we measure the TM using an orthonormal annular basis $\Phi$, instead of the conventional rectangular one. Following the approach of Ref.~\cite{Mastiani2021}, the system's circular aperture of radius $r_2 = 2.56$~mm is partitioned into $P=6$ concentric rings and a central circular region that fills the remaining aperture. Here, however, each ring is independently encoded using a Hadamard subbasis $H_p$ (Fig.~\ref{fig: 1}a), leading to a basis with a block-diagonal structure. The number of subbasis modes assigned to each ring is chosen proportional to its circumference, such that the outer rings contain more spatial modes than the inner ones. Consequently, the total number of input modes is $N=N_0+\sum_{p=1}^{P} N_p$=132, where $N_0=4$ and $N_p = \left[32,32,32,16,8,8\right]$ denotes the number of Hadamard vectors assigned to the central region and the $p$th ring (ordered from outermost to innermost), respectively. This construction allows every measured optical field to be represented as a vector whose coefficients are naturally grouped according to the annular region to which they belong.
	
	The propagation-invariant range associated with these spectral regions can be estimated from their DOF, which characterizes the axial range over which a beam undergoes limited transverse spreading while maintaining its intensity above half of its maximum value. Within this basis construction, two illumination conditions provide useful cases for comparison. Uniform illumination of the full circular aperture produces a -like beam after a transforming lens of focal length $f$ (see Fig.~S2), with $\text{DOF}_\text{C}=2z_R$, where $z_R=5.6f^2/kr_2^2$ is the effective Rayleigh length and $k=2\pi/\lambda_0$ is the wavenumber. In contrast, uniform illumination of an individual ring produces a Bessel beam with $\text{DOF}_\text{B}=5.6f^2/k\overline{r}\Delta r$ \cite{Mahmoud2013}, where $\overline{r}$ and $\Delta r$ are the mean radius and radial width of the ring, respectively. For the secondary imaging arm, with $f=75$~mm and a $0.3\times$ pre-demagnification, the corresponding theoretical values are $z_R=5.34$~mm and $\text{DOF}_{\text{B},p}\approx[6,8,10,19,31,81]z_R$. Even when two ($p = [1,2]$) or three ($p=[1,2,3]$) neighboring outer rings are combined, the propagation ranges remain $2.6z_R$ and $3.4z_R$, respectively. We therefore seek LSIMs with spectra corresponding to these annular regions, as they are expected to support extended propagation invariance. To contrast them, we also consider delocalized SIMs (DSIMs), whose spectra are broadly distributed across the available rings, which are expected to exhibit Gaussian-like behavior.
	
	
	The TM ($T$) relates the incident field ($E^\text{in}$) defined at the SLM to the transmitted field ($E^\text{out}$) measured at CAM1 via
	\begin{equation}
		E^\text{out} = T E^\text{in}
	\end{equation}
	To reconstruct $T$, we probe the system with $N$ input modes corresponding to the columns of the basis $\Phi$ and measure $M=N$ output modes. The $n$th input mode is phase shifted three times and the output is interfered with a reference beam, yielding the output intensity vectors $I^\text{out}_{n\delta}$, where $\delta$ is the phase step. The output field vector is recovered from the three-step PSI using~\cite{Scheidt2023hadamard}
	\begin{equation}
		E^\text{out}_{n}=-\frac{1}{3}(I^\text{out}_{n2}+I^\text{out}_{n3}-2I^\text{out}_{n1}) +\frac{i}{\sqrt{3}}(I^\text{out}_{n3}-I^\text{out}_{n2}) 
	\end{equation}
	and, by arranging each one as columns, $T$ is obtained in the annular basis. Though, it can be still returned to a canonical representation by performing the transformation $T\Phi^\dagger$.
	
	This procedure is done both in air and with the diffuser in order to obtain their TMs, $T_\text{a}$ and $T_\text{s}$, respectively.	Figs.~\ref{fig: 1}(b--d) show the measured TMs for the three scattering samples. As the scattering strength increases, energy progressively spreads outside the block-diagonal structure defined by the annular basis, indicating increasing modal crosstalk. In the weak diffuser, the transmitted modes remain largely confined to their original rings, whereas the moderate and strong diffusers increasingly couple light between different annular regions. A statistical characterization of the measured TMs is detailed in Sec.~S3 of the Supplemental Material.
	
	
	Once $T_\text{a}$ and $T_\text{s}$ have been measured, SIMs are obtained by solving the eigenvalue problem \cite{Pai2021}
	\begin{equation}
		T_\text{a}^{\dagger}T_\text{s} E_\nu = \gamma_\nu E_\nu
		\label{eq: sim_eq}
	\end{equation}
	where $T_\text{a}$ is assumed unitary, $E_\nu$ denotes the $\nu$th SIM and $\gamma_\nu$ is the complex transmission coefficient that accounts for the attenuation and phase shift introduced by the scattering medium. The calculated SIMs are experimentally generated using complex-amplitude modulation at the SLM following the method of Ref.~\cite{Bolduc2013}. 
	
	
	The measured SIMs for the weak, moderate, and strong scattering samples are shown in Figs.~\ref{fig: 2}(a--d). Despite the reduced dimension of the measured TMs ($N=132$), the reconstructed modes exhibit high fidelities (Eq.~S1), with several modes exceeding $0.9$ in all three scattering regimes (Figs.~S5). For the weak diffuser, the fidelities remain nearly constant across the complete set of SIMs, whereas the moderate and strong diffusers exhibit a gradual decrease with increasing mode index. This degradation is consistent with the combined effects of nonuniform illumination, which makes the measured free-space TM deviate from perfect unitarity \cite{Pai2021}, and incomplete control of the optical degrees of freedom resulting from the reduced-dimensional TM \cite{Goetschy2013,Yu2013,Rotter2017}. 
	
	A distinct evolution is observed in the SIM eigenvalue phases and their spectral localization. Although the moderate and strong diffusers exhibit similar TM statistics (Fig.~S4), their eigenvalue phase distributions differ substantially (Fig.~\ref{fig: 2}e), since the moderate diffuser retains partial phase correlations, whereas the strong diffuser approaches an almost uniform phase distribution. Annular localization also becomes progressively more difficult to preserve as scattering increases. For the weak diffuser, we identify $75$ LSIMs among the $128$ SIMs associated with the annular regions, of which $24$ are localized in $p=1$, $36$ in $p=[1,2]$, and $15$ in $p=[1,2,3]$. No LSIMs are initially identified for either the moderate or strong diffuser. To reveal weaker localization that may be obscured by low-intensity spectral components, we discard components below $20\%$ of its maximum intensity. After filtering, $25$ LSIMs are recovered for the moderate diffuser, with $16$ localized in $p=1$, $2$ in $p=[1,2]$, and $7$ in $p=[1,2,3]$, while none are recovered for the strong diffuser. The progressive disappearance of LSIMs highlights the sensitivity of annular localization to the modal crosstalk induced by scattering.
	
	To evaluate the SIMs propagation, their measured intensity evolution is compared with that of two beams: the Gaussian-like beam produced by uniformly illuminating the full circular aperture and the Bessel beam generated by uniformly illuminating ring $p=1$. The propagation is analyzed from the focal plane of lens $L_5$ up to approximately $5z_R$. Although the DOF of these beams can be determined in air, the speckle introduced by scattering makes its estimation unreliable after transmission through the samples. We therefore quantify preservation of the transverse intensity pattern using the Pearson correlation coefficient (PCC), taking the intensity distribution measured in air at the focal plane, $z_f=0$, as the common reference. We additionally define the correlation distance, $z_{0.5}$, as the distance at which the PCC decreases to $0.5$. In air, the Gaussian-like and Bessel beams exhibit $z_{0.5}=1.7z_R$ and $4.6z_R$, respectively. We also consider the characteristic planes $2z_R$ and $4z_R$, which lie outside the Gaussian-like and Bessel DOF regions, respectively. The propagation evolution for the weak and moderate scattering samples is shown in Fig.~\ref{fig: 3}, while the strong scattering results are presented in Fig.~S6.
	
	The Gaussian-like and Bessel beams exhibit a rapid loss of propagation robustness as the scattering strength increases. Under weak scattering, their evolution remains relatively close to that observed in air. In the moderate regime, their correlation distances decrease to only $z_{0.5}=0.12z_R$ for the Gaussian-like beam and $0.21z_R$ for the Bessel beam, while in the strong regime they reach $0.10z_R$ and $0.19z_R$, respectively. Thus, the extended propagation range of the Bessel beam observed in air is largely suppressed by scattering, and both beams decorrelate within a small fraction of $z_R$ under moderate and strong scattering.
	
	The LSIMs show a markedly different behavior. Under weak scattering, the LSIM shown in Fig.~\ref{fig: 3}b retains its main intensity features and exhibits Bessel-like propagation, reaching $z_{0.5}=4.2z_R$. This value is only slightly shorter than the obtained for the same LSIM in air and for the Bessel beam after the diffuser. Its high correlation is also maintained at $2z_R$ and $4z_R$ (Fig.~\ref{fig: 3}g), showing that scattering and propagation invariance can coexist when modal crosstalk remains low. In the moderate regime, strict nondiffracting propagation is no longer preserved, yet the filtered LSIM still reaches $z_{0.5}=0.63z_R$, approximately three times the correlation distance of the Bessel beam and more than five times that of the Gaussian-like beam. No LSIMs are identified in the strong scattering regime, consistent with the loss of annular localization at high modal crosstalk.
	
	On the other hand, since the angular spectra of DSIMs extend over a broad region, these modes undergo continuous structural changes even in air and therefore do not exhibit propagation invariance. Under weak scattering, the DSIM in Fig.~\ref{fig: 3}e has the same correlation distance as the Gaussian-like beam, nevertheless, its correlation decays more slowly at longer distances, remaining approximately four times higher at $4z_R$. The relative advantage of the DSIM becomes more pronounced as scattering increases. For the moderate diffuser, it reaches $z_{0.5}=0.58z_R$, nearly three times the Bessel correlation distance and about five times that of the Gaussian-like beam. This advantage persists for the strong diffuser, where the DSIM reaches $z_{0.5}=0.32z_R$, retaining its correlation over $1.7$ and $3.2$ times that of the Bessel and Gaussian-like beams. Although these modes are not propagation invariant in the strict sense, they retain their transverse intensity structure more effectively under moderate and strong scattering.
	
	In conclusion, we have shown that SIMs can acquire propagation-invariant behavior when measured in an annular basis and modal crosstalk remains sufficiently low. We identify $75$ LSIMs for the weak diffuser, $25$ for the moderate diffuser after spectral filtering, and none for the strong diffuser, demonstrating the progressive loss of annular localization with increasing scattering. Under weak scattering, a representative LSIM retains Bessel-like propagation, with a correlation distance $z_{0.5}$ comparable to that of the Bessel beam. Although this nondiffracting behavior is lost as scattering increases, SIMs remain more robust during propagation than the Gaussian-like and Bessel beams.
	
	\begin{backmatter}
		\bmsection{Funding} We acknowledge support from the UNAM-PAPIIT grants IG101425; CTIC-LANMAC; SECIHTI LN-299057 and CF-2023-I-657. J.C.O.T acknowledge the SECIHTI graduate studies funding 1187782.
		
		\bmsection{Acknowledgment} Thanks to Cristian Mojica Casique for technical support and Jos\'e Rangel for machine shop work.
		
		\bmsection{Disclosures} The authors declare no conflicts of interest.
		
		\bmsection{Data availability} Data underlying the results presented in this paper are not publicly available at this time but may be obtained from the authors upon reasonable request.
		
		\bmsection{Supplemental document} See Supplement 1 for supporting content.
		
	\end{backmatter}
	
	\bibliography{Refs.bib}
	
\end{document}